\documentstyle[epsf,multicol,aps]{revtex}

\def\l{\langle}
\def\r{\rangle}

\begin{document}
\draft
\title{
Three-dimensional antiferromagnetic $q$-state Potts models:
application of the Wang-Landau algorithm
}

\author{Chiaki Yamaguchi\cite{yama} and Yutaka Okabe\cite{okabe}}
\address{
Department of Physics, Tokyo Metropolitan University,
Hachioji, Tokyo 192-0397, Japan
}

\date{Received \today}

\maketitle

\begin{abstract}
We apply a newly proposed Monte Carlo method, the Wang-Landau algorithm, 
to the study of the three-dimensional antiferromagnetic 
$q$-state Potts models on a simple cubic lattice.  
We systematically study the phase transition 
of the models with $q$=3, 4, 5 and 6.  
We obtain the finite-temperature phase transition for $q$= 3 and 4, 
whereas the transition temperature is down to zero for $q$=5. 
For $q$=6 there exists no order for all the temperatures. 
We also study the ground-state properties.
The size-dependence of the ground-state entropy is 
investigated.  We find that the ground-state entropy 
is larger than the contribution from the typical configurations 
of the broken-sublattice-symmetry state for $q=3$. 
The same situations are found for $q$ = 4, 5 and 6. 

\end{abstract}


\begin{multicols}{2}
\narrowtext

\section{Introduction}

The $q$-state Potts model has various interesting properties 
to study \cite{Potts,Kihara,Wu82}.  The order of the phase 
transition of the Potts model depends on the spatial dimensionality 
and the number of states, $q$.  The phase transitions of 
the antiferromagnetic (AF) Potts models are more complex than 
those of the ferromagnetic Potts models. 
Here, we focus on the three-dimensional (3D) AF 
$q$-state Potts models.  Banavar {\it et al.} \cite{BGJ80} 
studied the AF 3- and 4-state Potts models 
by use of the Monte Carlo simulation.  Ono \cite{Ono86} pointed out 
an appropriate choice of the vector order parameter 
for the AF Potts models on a bipartite lattice.  
The phase transition of the AF 3-state Potts model 
is considered to belong to the $XY$ universality class \cite{WSK89,GH94}.  
The low-temperature phase of the AF 3-state Potts model 
was shown to be the broken-sublattice-symmetry (BSS) state, 
but it is still an open question whether there exists the rotationally 
symmetric state or not below the 2nd order phase transition point 
of the $XY$ universality class \cite{KS95,HSK96,Rahman98,Oshikawa00}. 
Compared to the AF 3-state Potts model, there have been 
not so many systematic studies on the AF 4-state and 
higher-state Potts models in three dimensions.
Recently, the AF 4-state Potts model was studied by Itakura \cite{Ita99};  
there is a finite-temperature phase transition, and 
the critical phenomena may belong to the Heisenberg universality 
class if the phase transition is of 2nd order.  However,  
a possibility of the 1st order transition was also argued \cite{Ita99}.
Another source of interest in AF $q$-state Potts 
models is that it has the nonzero ground-state entropy 
without frustration.  Nonzero ground-state entropy, $S_0 \ne 0$, 
is an important subject in statistical mechanics.  
One physical example is provided by ice. 
The AF Potts model is served as a useful model 
for the study of the nonzero ground-state entropy,
and the nonzero ground-state entropy of the two-dimensional (2D)
AF Potts model was extensively studied 
by Shrock and Tsai \cite{Shrock}.

The Monte Carlo simulation is regarded as a standard tool 
for studying statistical mechanical properties \cite{Landau00}. 
The Monte Carlo study using a conventional Metropolis algorithm 
sometimes suffers from the problem of slow dynamics, or the 
long time scale problem.  In the study of the phase transition,
the long time scale due to the critical slowing down causes 
the problem of slow equilibration.
For the simulational study of the AF Potts models, 
the cluster algorithm \cite{sw87}
has been used \cite{WSK89,GH94,HSK96,Rahman98,Ita99} 
to overcome the critical slowing down.  
As an extension of the cluster algorithm, Tomita and 
Okabe \cite{to01} recently proposed a new cluster algorithm 
of tuning the critical point automatically, 
and applied it to the study of 2D Potts models. 
The extended ensemble method is another type of attempts 
to overcome the slow dynamics.  The multicanonical 
method \cite{berg91,Lee93}, 
the simulated tempering \cite{marinari92}, 
the exchange Monte Carlo method 
\cite{hukushima96} (or the multiple Markov chain method \cite{tesi96}), 
the broad histogram method \cite{oliveira98}, 
and the flat histogram method \cite{wang98} are examples of 
the extended ensemble method.  Quite recently, Wang and 
Landau \cite{wl01} proposed an efficient algorithm to calculate 
the density of states with high accuracy.  In the present paper 
we use the Wang-Landau algorithm \cite{wl01} to study 
the 3D AF Potts models.  We study not only the phase transition of 
the AF Potts models but also the ground-state properties. 
The Wang-Landau algorithm is quite effective for this purpose 
because we can calculate the density of states with high accuracy. 
We make a random walk in the whole energy space, and the algorithm 
is appropriate for calculating the ground-state entropy. 

We organize the rest of the paper as follows. 
In Sec. II, we describe the model and the vector order 
parameter for the AF Potts model.  We also briefly explain 
the simulation method, the Wang-Landau algorithm.  
In Sec. III, we study the phase transition of the AF 
3-, 4-, 5- and 6-state Potts models on a simple cubic lattice.
The ground-state properties of the 3-, 4-, 5- and 6-state Potts 
models are studied in Sec. IV.
The summary and discussions are given in Sec. V. 

\section{Model and Simulation method}

We deal with the AF Potts model defined by the Hamiltonian
\begin{equation}
  {\cal H} = J \sum_{<i,j>} \delta_{\sigma_i,\sigma_j}, \quad (J>0),
\label{Hamiltonian}
\end{equation}
where the Potts variable $\sigma_i$ takes the value $1, 2, \cdots, q$,
and the summation is taken over the nearest-neighbor pairs of sites 
on a simple cubic lattice.

The ferromagnetic 3-state Potts model can be transformed 
into the $Z_3$ clock model, and the order parameter is 
well represented by a two-dimensional vector. 
For the AF Potts model on a bipartite lattice, the sublattice 
structure should be taken into account \cite{Ono86,Okabe95}; 
that is, the staggered magnetization will be treated.
Using three components of the staggered magnetization 
\begin{equation}
 m_{i} = \Big(  \sum_{j \in A} \delta_{\sigma_j,i}
            - \sum_{j \in B} \delta_{\sigma_j,i} \Big)\Big/N ,
\label{order}
\end{equation}
where $A$ and $B$ denote two sublattices, we define the vector 
order parameter as follows:
\begin{equation}
 {\bf M} \equiv m_1 {\bf e}_1 + m_2 {\bf e}_2 + m_3 {\bf e}_3.
\end{equation}
Here, ${\bf e}_i$ denotes the unit vectors in two dimensions 
directing 120 degrees apart from each other, and 
\begin{equation}
 {\bf e}_1 + {\bf e}_2 + {\bf e}_3 = 0.
\label{sum}
\end{equation}
Then, the square of the vector order parameter simply
becomes
\begin{equation}
  {\bf M}^2 = \frac{3}{2} (m_1^2+m_2^2+m_3^2).
\label{vec_order_2}
\end{equation}
We can extend the above argument to higher-state Potts models
\cite{Ono86}. In general the order parameter of $q$-state 
Potts model is well described by the $(q-1)$ dimensional vector 
order parameter.  We consider the unit vectors in $(q-1)$ dimensions 
pointing to $q$ directions as in the case of the 3-state 
Potts model; 
the sum of the unit vectors is set to be zero as in Eq.~(\ref{sum}).
Using each component of the staggered magnetization, 
Eq.~(\ref{order}), we can obtain the generalized expression 
for Eq.~(\ref{vec_order_2}),
\begin{equation}
  {\bf M}^2 = \frac{q}{q-1} (m_1^2+m_2^2+ \cdots + m_q^2).
\end{equation}
We should note that for the 3-state Potts model 
the vector order parameter space spans 
a hexagonal region in a two-dimensional space, and the 
maximum value of $|{\bf M}|$ is $\sqrt{3}/2$.
For the 4-state Potts model the order parameter 
takes the values within the area shown in Fig.~3 of 
Ref.~\cite{Ita99}, and the maximum value of 
$|{\bf M}|$ is $\sqrt{2/3}$.

We briefly describe the Wang-Landau algorithm \cite{wl01}.
This algorithm is similar to Lee's version of the multicanonical 
method (entropic sampling) \cite{Lee93}, 
the broad histogram method \cite{oliveira98}, and the flat 
histogram method \cite{wang98}; but the Wang-Landau algorithm 
has advantage that it can estimate the density of states 
efficiently even for large systems. 
The idea of the Wang-Landau algorithm \cite{wl01} is that we make 
a random walk in energy space based on the transition probability 
from energy level $E_{1}$ to $E_{2}$; 
\begin{equation}
p(E_{1}\rightarrow E_{2})=\min \Big[ \frac{g(E_{1})}{g(E_{2})},1 \Big], 
\end{equation}
where $g(E)$ is the density of states.  Since the exact form of $g(E)$ 
is not known {\it a priori}, we determine $g(E)$ iteratively; 
$g(E)$ is modified by
\begin{equation}
\ln g(E) \rightarrow \ln g(E) + \ln f_i,
\end{equation}
every time the state is visited.  The modification factor $f_i$ 
is gradually reduced to 1 by checking the ``flatness" 
of the energy histogram; the histogram for all possible $E$ 
is not less than some value of the average histogram, say, 80\%.

We simulate the AF $q$-state Potts model 
($q$=3, 4, 5, 6) on a simple cubic lattice 
by using the Wang-Landau algorithm \cite{wl01}. 
We impose the periodic boundary conditions and 
the linear sizes are $L$ =8, 10, 12, 14 and 16.
For the modification factor $f_i$, we start with 
$f_0 = e^k$ with $k$=1 or some positive integer,
and $f_{i+1} = \sqrt{f_i}$; the final value of $f_i$ is 
chosen as $10^{-8}$, which is the same as Ref.~\cite{wl01}. 
We calculate the density of states $g(E)$, and measure 
the physical quantities of interest as a function of $E$. 
Then, the canonical average of the physical quantity $Q$ 
at the inverse temperature $\beta=1/k_{B}T$ is 
calculated thorough the standard relation
\begin{equation}
 \l Q \r_{\beta} = \frac{\int Q(E) g(E) e^{-\beta E} \ dE}
                {\int g(E) e^{-\beta E} \ dE}.
\end{equation}
In the actual calculation, the relative density of states,
$g(E_1)/g(E_2)$, is directly obtained.  In terms of the entropy 
(in units of $k_B$), $S(E)=\ln g(E)$, the entropy difference, 
$S(E_1)-S(E_2)$, is directly measured.  
Imposing the constraint 
\begin{equation}
  \sum_E g(E) = q^N, 
\end{equation}
we can determine the absolute value of $g(E)$.  
Here, $N(=L^3)$ is the number of lattice sites, that is, 
the number of Potts spins. 
For the AF Potts model on a simple cubic lattice, 
Eq.~(\ref{Hamiltonian}), the energy $E$ takes the 
value from 0 to $3N$ in units of $J$.  The state with 
the highest energy $E=3N$ is nothing but the ferromagnetic 
ground state, and the degeneracy of the ferromagnetic ground states 
is $q$.  Therefore, we may check the accuracy of the calculation 
by confirming $g(3N) = q$.

\section{Phase Transitions of AF Potts model}

\begin{figure}
\epsfxsize=\linewidth 
\centerline{\epsfbox{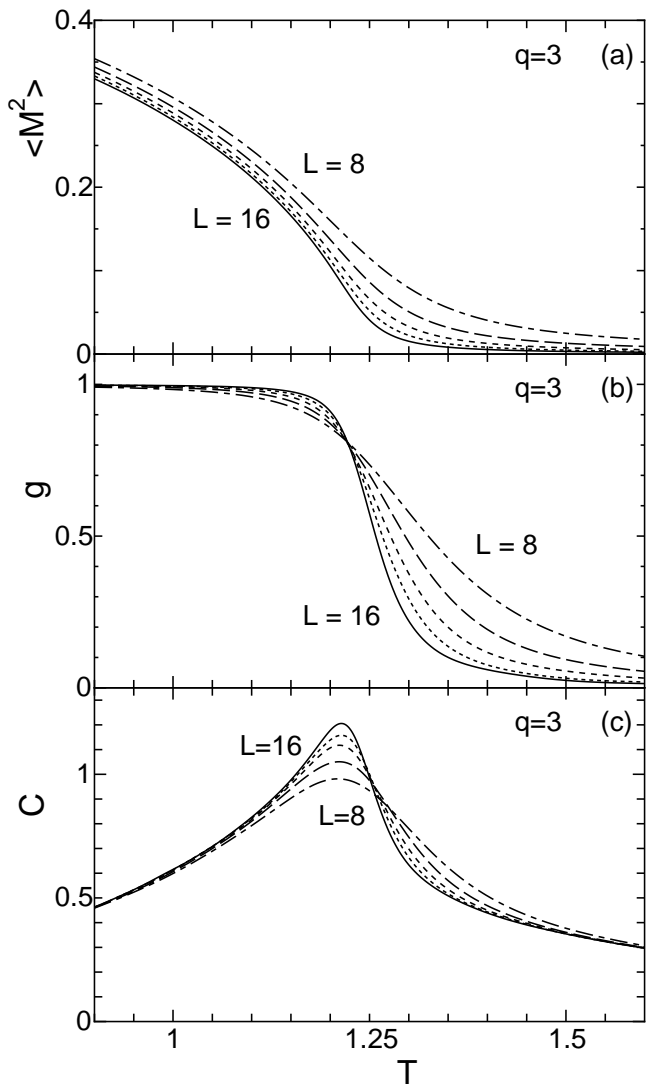}}
\caption{
Temperature dependence of the squared order parameter (a), 
the Binder parameter (b), and the specific heat (c)
for the 3D AF 3-state Potts model. 
The linear system sizes are $L$ = 8, 10, 12, 14 and 16.
The dot-dashed, dashed, short-dashed, dotted and solid lines 
are used for $L$=8, 10, 12, 14 and 16, respectively.
The temperature is represented in units of $J/k_B$.
}
\label{fig3}
\end{figure}

First, we study the phase transition of the AF Potts model 
with $q$=3, 4, 5 and 6.  Let us start with showing the 
data for $q=3$ in order to make a comparison with higher-state 
models, although the phase transition 
of the 3-state Potts model has been studied extensively 
\cite{WSK89,GH94,KS95,HSK96,Rahman98}.  The temperature dependence 
of $\l {\bf M}^2 \r$ for $q$=3 is shown in Fig.~\ref{fig3}(a). 
The data for $L$=8, 10, 12, 14 and 16 are plotted by 
the dot-dashed, dashed, short-dashed, dotted and 
solid lines, respectively.
The temperature is represented in units of $J/k_B$. 
All the measurements are done for 4 independent runs, 
and the average is taken over 4 samples.
The normalized fourth-order cumulant of the magnetization, 
the Binder parameter
\cite{Binder81}
\begin{equation}
  g = \frac{q+1}{2} \Big(1 - \frac{q-1}{q+1} 
  \frac{\l {\bf M}^4 \r}{\l {\bf M}^2 \r^2} \Big),
\label{Binder}
\end{equation}
is plotted in Fig.~\ref{fig3}(b).  
The normalization factors in Eq.~(\ref{Binder}) 
are chosen such that 
\begin{equation}
 g \rightarrow   \left\{
                 \begin{array}{lll}
                 1 \quad {\rm for} \quad T=0 \\
                 0 \quad {\rm for} \quad T=\infty
                 \end{array}
                 \right.
\end{equation}
by taking account of the $(q-1)$ dimensional vector structure of 
the order parameter.
The definition of Eq.~(\ref{Binder}) becomes a usual one 
for the scalar order parameter ($q=2$). 
Since the prefactors of the $L$ 
dependence in the finite-size scaling equations
are canceled out, one may determine the critical temperature $T_c$
from the crossing point of the data of temperature dependence 
for different sizes as far as the corrections to finite-size 
scaling are negligible.  
We also plot the specific heat in Fig.~\ref{fig3}(c).  
The specific heat peak becomes sharper when the system size 
is larger.  From Figs.~\ref{fig3}(a), \ref{fig3}(b) and \ref{fig3}(c),
we find a clear phase transition at a finite temperature.

\begin{figure}
\epsfxsize=\linewidth 
\centerline{\epsfbox{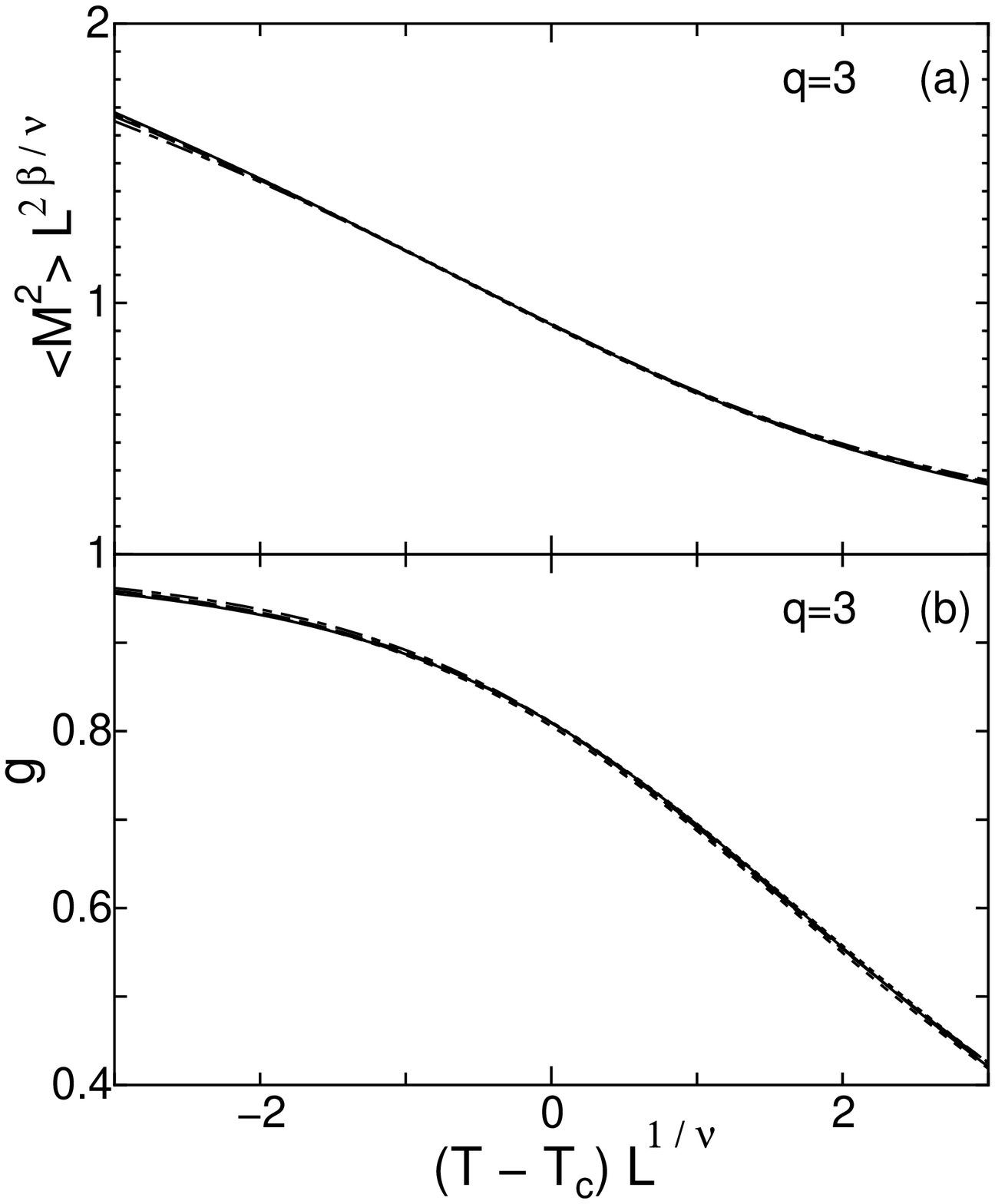}}
\caption{
Scaling plots of the order parameter (a) and the
Binder parameter (b) for the 3D AF 3-state Potts model. 
The linear system sizes are $L$ = 8, 10, 12, 14 and 16.
The estimated values for the critical 
temperature and the critical exponents are
$T_c$ = 1.222, $1/\nu$= 1.52 and $\beta/\nu$=0.46; 
}
\label{fig3_scale}
\end{figure}

The finite-size scaling plots of the order parameter,
and the Binder parameter 
\begin{eqnarray}
  \l {\bf M}^2 \r &=& L^{-2\beta/\nu} f\Big( (T-T_c) L^{1/\nu} \Big), \\
  g &=& g\Big( (T-T_c) L^{1/\nu} \Big), 
\end{eqnarray}
are given in Figs.~\ref{fig3_scale}(a) and \ref{fig3_scale}(b), 
respectively.  
The estimated values for the critical 
temperature and the critical exponents are
$T_c$ = 1.222(4), $1/\nu$= 1.52(4) and $\beta/\nu$=0.46(5). 
Here, the number in the parentheses denotes the uncertainty 
in the last digit. 
The estimated values are consistent with the previous studies 
\cite{WSK89,GH94,KS95,HSK96}, and the obtained exponents 
are close to those of the 3D $XY$ model. 
We have given the estimate using the finite-size scaling analysis
to check the consistency of the calculation. 
Precisely speaking, due to small system sizes, 
our estimate of $T_c$ is a little bit 
lower than the accurate estimate \cite{WSK89,GH94}, 
which results in a little bit smaller $\beta/\nu$. 
To discuss more accurate estimates of $T_c$ and the critical 
exponents, calculations with larger system sizes are 
preferable.

\begin{figure}
\epsfxsize=\linewidth 
\centerline{\epsfbox{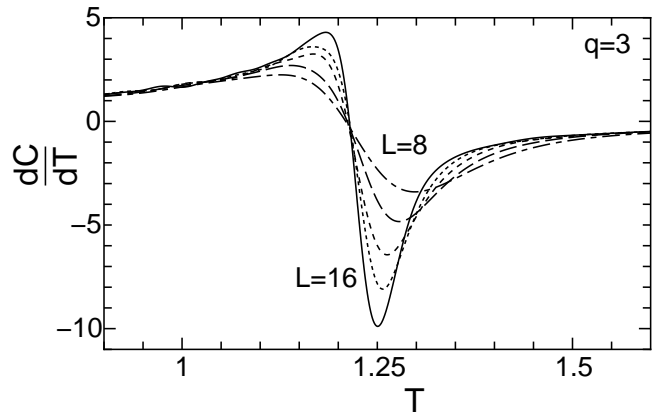}}
\caption{
Temperature derivative of the specific heat 
for the 3D AF 3-state Potts model. 
The linear system sizes are $L$ = 8, 10, 12, 14 and 16.
}
\label{fig3_cd}
\end{figure}

It is not easy to estimate the specific heat exponent $\alpha$ 
from the specific heat data using the finite-size scaling analysis, 
if $\alpha$ is negative \cite{WSK89}. 
We may use the temperature derivative of the specific heat, 
which is singular at $T_c$ even if $\alpha$ is negative. 
We plot the temperature derivative of the specific heat 
in Fig.~\ref{fig3_cd}.  
Since we directly calculate the density of states by using the 
Wang-Landau algorithm, it is easy to compute the temperature 
derivative of the specific heat from the moments of energy.  
Using the finite-size scaling relation
\begin{equation}
  \frac{d C}{d T} = L^{(\alpha-1)/\nu} f\Big( (T-T_c) L^{1/\nu} \Big), 
\end{equation}
we estimate the exponent $\alpha$ as -0.04(6).  
Although our estimate has a relatively large error bar 
due to the small system size, 
our result suggests that $\alpha$ is negative.

\begin{figure}
\epsfxsize=\linewidth 
\centerline{\epsfbox{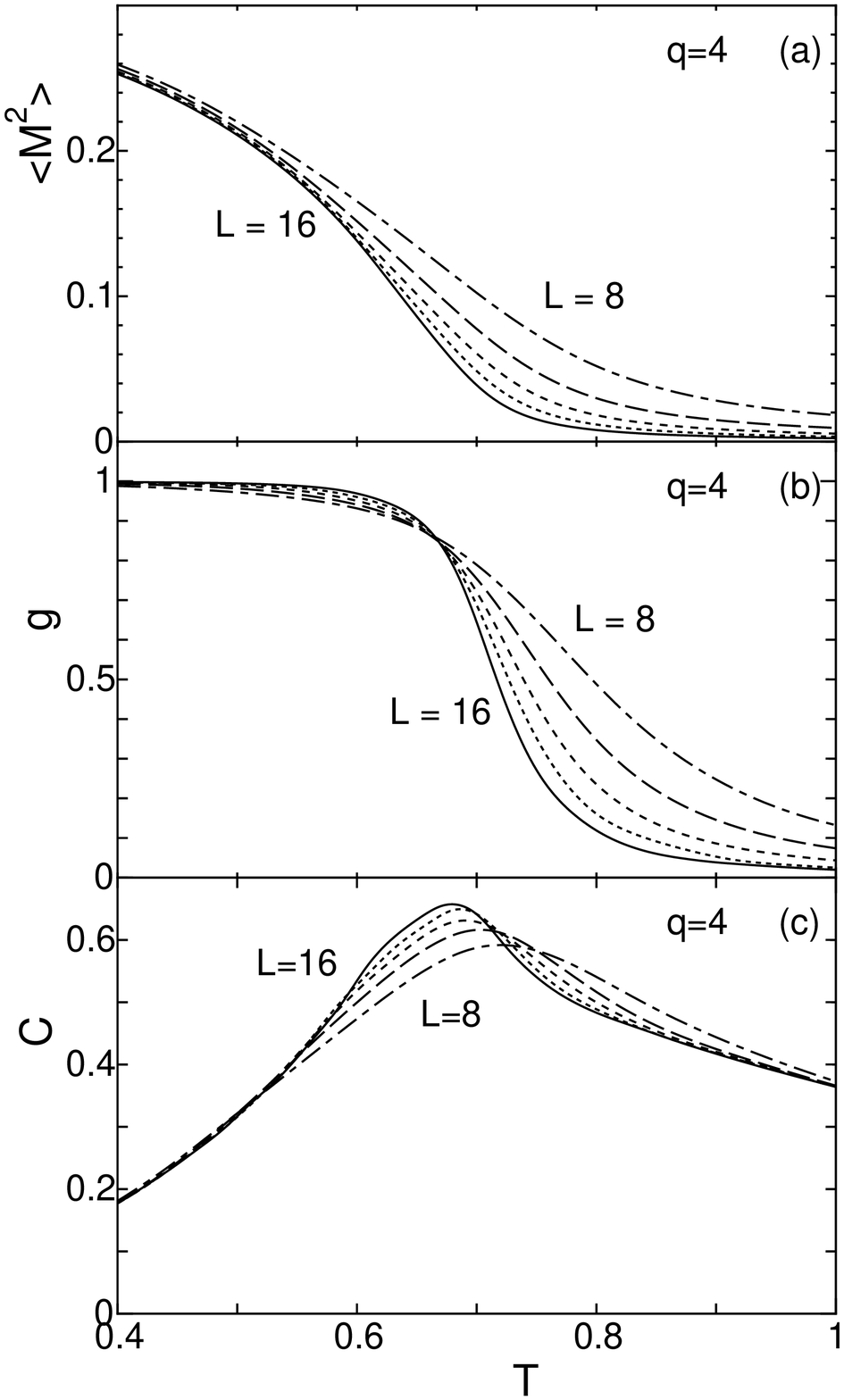}}
\caption{
Temperature dependence of the squared order parameter (a),  
the Binder parameter (b), and the specific heat (c)
for the 3D AF 4-state Potts model. 
The linear system sizes are $L$ = 8, 10, 12, 14 and 16.
The temperature is represented in units of $J/k_B$.
}
\label{fig4}
\end{figure}

\begin{figure}
\epsfxsize=\linewidth 
\centerline{\epsfbox{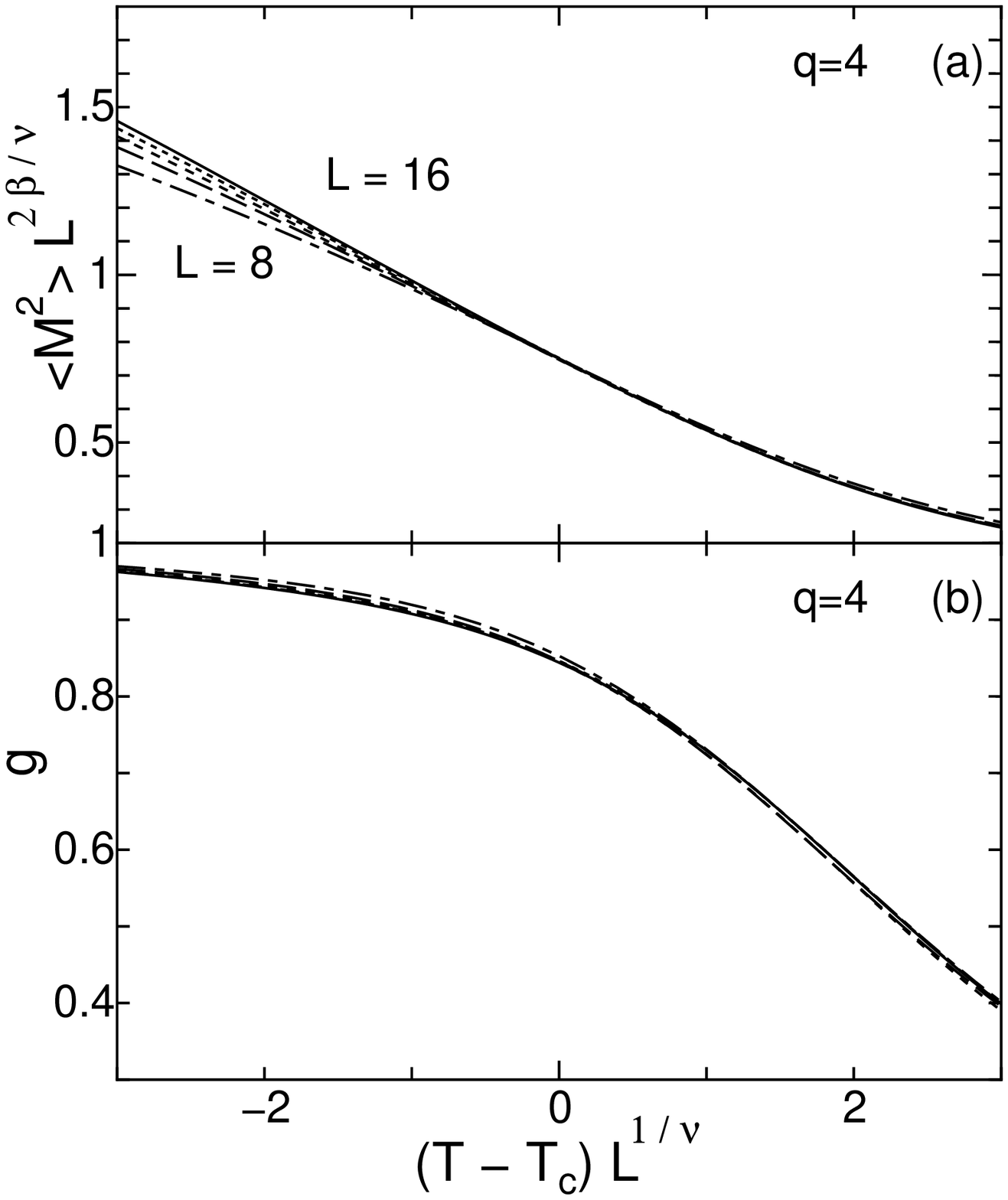}}
\caption{
Scaling plots of the order parameter (a) and 
the Binder parameter (b) for the 3D AF 4-state Potts model. 
The linear system sizes are $L$ = 8, 10, 12, 14 and 16.
The estimated values for the critical 
temperature and the critical exponents are
$T_c$ = 0.669, $1/\nu$= 1.41 and $\beta/\nu$=0.44; 
}
\label{fig4_scale}
\end{figure}

Next turn to the 4-state Potts model.  We plot the temperature dependence of 
the order parameter and the Binder parameter in Fig.~\ref{fig4}(a) and 
\ref{fig4}(b), respectively.  We also show the specific heat 
in Fig.~\ref{fig4}(c).  
We here make a comment on the statistical errors for the 
estimate of the density of states.  They becomes larger
for higher $q$ and larger $L$.  The errors for the specific heat 
curve for larger $L$ are larger than the thickness of the curve 
in Fig.~\ref{fig4}(c). 
From Fig.~\ref{fig4} we find that there exits a clear 
finite-temperature phase transition. 
The finite-size scaling plots of the order parameter 
and the Binder parameter are given in Figs.~\ref{fig4_scale}(a) 
and \ref{fig4_scale}(b), respectively; 
we estimate $T_c$, $1/\nu$ and $\beta/\nu$ 
as 0.669(4), 1.41(4) and 0.44(6).  They are compatible 
with the previous study \cite{Ita99}, and the obtained exponents 
are close to those of the 3D Heisenberg model.  
Precisely, our estimate of $T_c$ is a little bit lower than the accurate 
estimate \cite{Ita99}; the situation is the same as the case 
of $q=3$.

\begin{figure}
\epsfxsize=\linewidth 
\centerline{\epsfbox{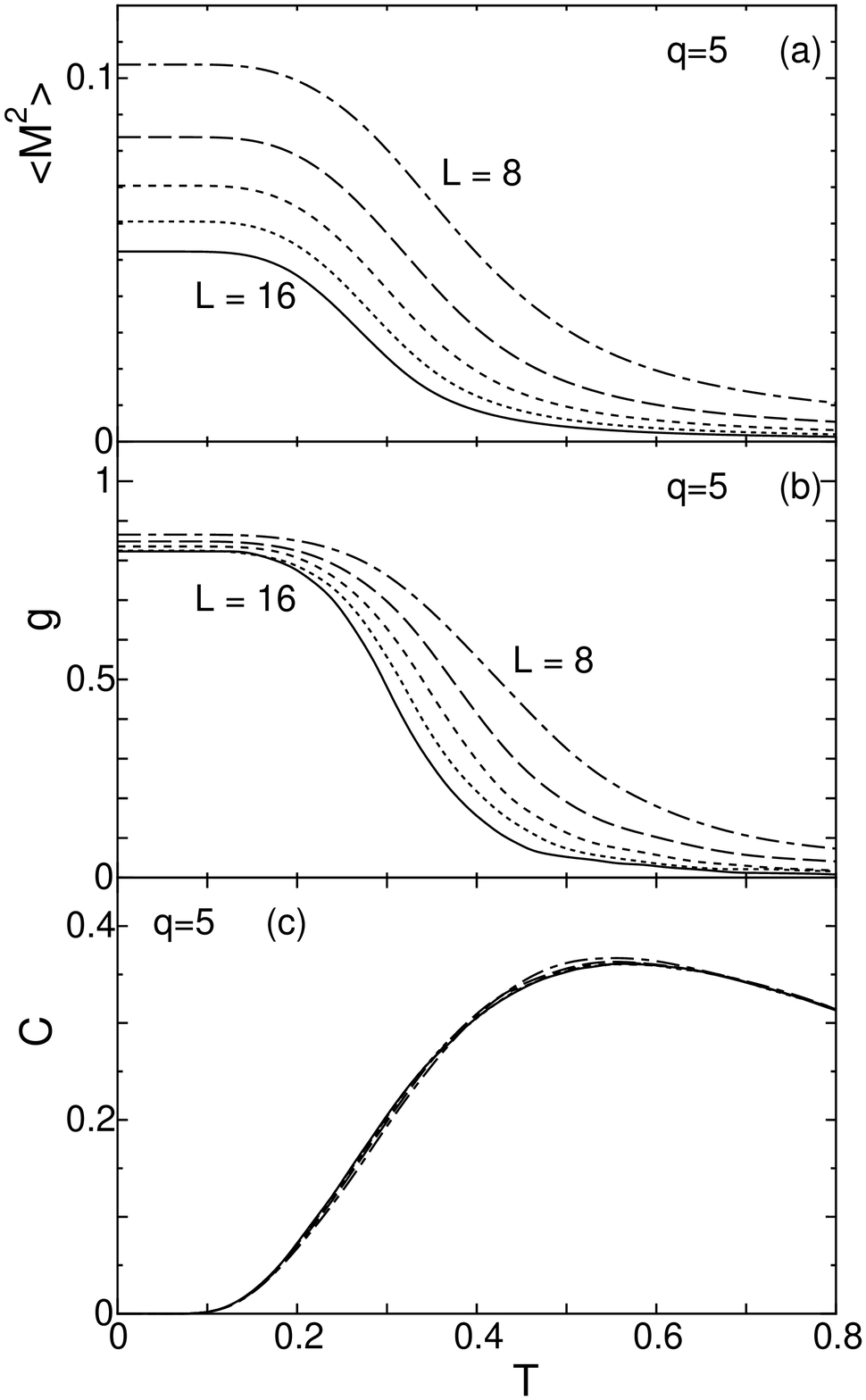}}
\caption{
Temperature dependence of the squared order parameter (a),
the Binder parameter (b), and the specific heat (c)
for the 3D AF 5-state Potts model. 
The linear system sizes are $L$ = 8, 10, 12, 14 and 16.
The temperature is represented in units of $J/k_B$.
}
\label{fig5}
\end{figure}

The order parameter, the Binder parameter and the specific heat 
for the 5-state Potts model are given in Figs.~\ref{fig5} (a), 
\ref{fig5}(b) and \ref{fig5}(c), 
respectively.  There is no anomaly in the specific heat.  
There is no crossing in the Binder parameter at finite temperatures. 
The critical temperature is down to zero.  
However, the value of the Binder parameter at $T=0$ is finite, 
and it may become constant for large enough $L$; 
it is not clear whether $T=0$ is 
critical or not.  To determine this point, 
more elaborate study with larger sizes are necessary.

\begin{figure}
\epsfxsize=\linewidth 
\centerline{\epsfbox{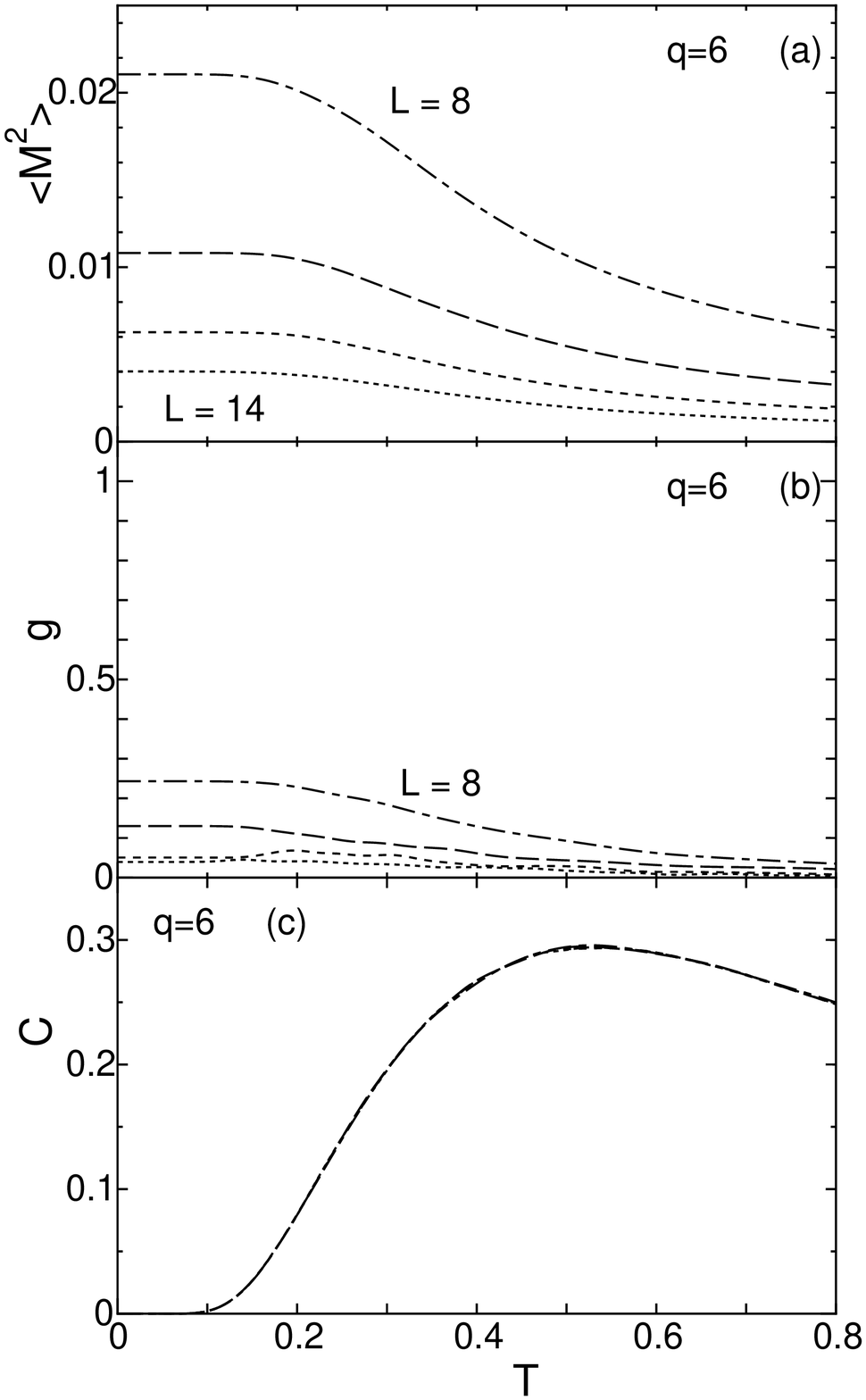}}
\caption{
Temperature dependence of the squared order parameter (a),
the Binder parameter (b), and the specific heat (c)
for the 3D AF 6-state Potts model. 
The linear system sizes are $L$ = 8, 10, 12 and 14.
The temperature is represented in units of $J/k_B$.
}
\label{fig6}
\end{figure}

We show the order parameter, the Binder parameter and the specific heat 
for the 6-state Potts model in Figs.~\ref{fig6}(a), 
\ref{fig6}(b) and \ref{fig6}(c), respectively.  In this case, 
the maximum linear size is $L=14$. 
There is no anomaly in the specific heat.  
The value of the Binder parameter at $T=0$ is very small. 
This means that the distribution of the order parameter is Gaussian; 
in other words, the system is disordered.  We may conclude that 
there is no order even at $T=0$.  

\section{Ground-state Properties}

In this section, we focus on the ground-state properties 
of the AF Potts models.  First, we consider 
the ground-state entropy per spin, $S_0/N$; 
the entropy (in units of $k_B$) is calculated through 
$S_0=\ln g(0)$

\begin{figure}
\epsfxsize=\linewidth 
\centerline{\epsfbox{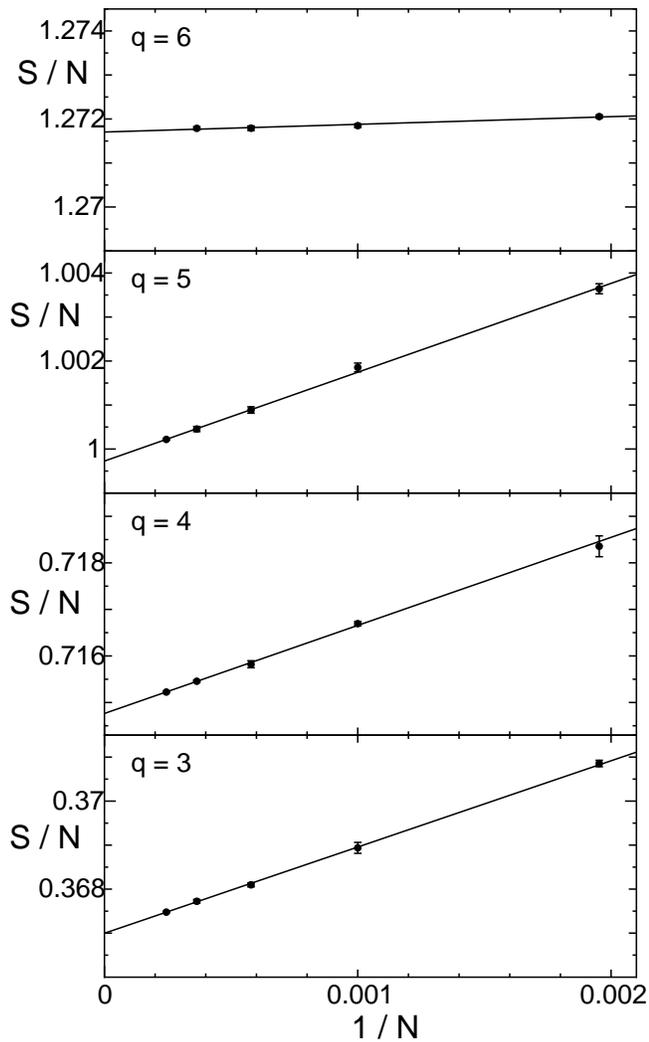}}
\caption{
Size dependence of the ground-state entropy per spin 
for the 3D AF $q$-state Potts models; $q$=3, 4, 5, and 6. 
The linear system sizes are $L$ = 8, 10, 12, 14 and 16 for 
$q$=3 to 5 and $L$ = 8, 10, 12 and 14 for $q$=6; $N=L^3$. 
}
\label{fig_ge}
\end{figure}

In Fig.~\ref{fig_ge}, we show the size dependence of 
the ground-state entropy per spin for the $q$-state 
AF Potts model on a simple cubic lattice.   
We plot $S_0/N$ as a function of $1/N$, and find a linear $1/N$ 
dependence.  Using the least square method, for $q=3$ 
we have 
\begin{equation}
 S_0/N = 0.3670(1) + 1.97(4) \times (1/N), \quad q=3, 
\label{entropy_3}
\end{equation}
where the number in the parentheses denotes the uncertainty 
in the last digit.  Wang {\it et al.} \cite{WSK89} estimated 
the ground-state entropy per spin ($ N \rightarrow \infty$) 
as 0.3673 from the data of $L$= 4 and 8.  Our estimate, 
0.3670, is a little bit smaller than their estimate. 
The low-temperature phase of the AF 3-state 
Potts model is the BSS state.
The typical configuration of the BSS state 
is that all the spins on the sublattice $A$ take one of the three states, 
and the spins on the sublattice $B$ take one of the other two states 
randomly.  Then, the lower bound for the ground-state entropy 
becomes $S_0/N \ge (\ln 2)/2 + \ln 6 \times (1/N) = 
0.3466 + 1.79 \times (1/N)$. 
Our estimate, Eq.~(\ref{entropy_3}), is, of course, 
larger than the lower bound. 
That is, there are many configurations other than 
the typical configurations of the BSS state.

From the linear $1/N$ dependence for the AF 4-state Potts model, 
we have 
\begin{equation}
 S_0/N = 0.7148(1) + 1.92(6) \times (1/N), \quad q=4.
\label{entropy_4}
\end{equation}
As far as we know, there has been no study on the estimate of 
the ground-state entropy for $q=4$ and higher $q$. 
For the low-temperature phase of the 4-state Potts model, 
the typical spin configuration is as follows:  
The spins on the sublattice $A$ take two of four states randomly, 
whereas those on the sublattice $B$ take the other two states. 
Then, the lower bound for the ground-state entropy 
becomes $S_0/N \ge \ln 2 + \ln 6 \times (1/N) = 0.6931 + 1.79 \times (1/N)$.
Our estimate, Eq.~(\ref{entropy_4}), is again 
larger than the lower bound.

From the size dependence of the ground-state entropy 
for the AF 5-state Potts model, which is also shown 
in Fig.~\ref{fig_ge}, we have 
\begin{equation}
 S_0/N = 0.9997(1) + 2.03(6) \times (1/N), \quad q=5.
\label{entropy_5}
\end{equation}
The main contribution to the low-temperature phase of 
the 5-state Potts model is as follows:
The spins on the sublattice $A$ take two of five states randomly, 
whereas those on the sublattice $B$ take the other three states. 
Then, the lower bound for the ground-state entropy 
becomes $S_0/N \ge (\ln 6)/2 + 2 \ln 5 \times (1/N) = 
0.8959 + 3.00 \times (1/N)$.
Our estimate, Eq.~(\ref{entropy_5}), is again 
larger than the lower bound.

Finally, for $q=6$ we have 
\begin{equation}
 S_0/N = 1.2717(1) + 0.17(4) \times (1/N), \quad q=6,
\label{entropy_6}
\end{equation}
by using the least square method. 
The main contribution of the 6-state Potts model is as follows:
The spins on the sublattice $A$ take three of six states randomly, 
whereas those on the sublattice $B$ take the other three states. 
Our estimate, Eq.~(\ref{entropy_6}), is again 
larger than the lower bound for the ground-state entropy; 
$S_0/N \ge \ln 3 + 2 \ln 5 \times (1/N) = 
1.0986 + 3.00 \times (1/N)$.

It is interesting to note that the ground-state entropy has 
a similar $1/N$ dependence for all $q=3,4,5$,
although there occurs the finite-temperature phase transition 
for $q=3,4$ and the zero-temperature phase transition for $q=5$. 
The size dependence of $q=6$ is similar to that for $q=3,4,5$.
But the slope of $1/N$ dependence for $q=6$ is smaller than that of 
others.  It is not clear whether this small difference is related 
to the existence of the antiferromagnetic order.  

We find that the ground-state entropy is larger than 
the contribution from the typical configurations 
of the BSS state.   In order to look into the 
ground-state properties more carefully, 
let us consider the proportion of visiting 
the typical BSS ground states among all the ground states.  
As an example, we treat the case of $q=3$.  
Since this proportion becomes very small for larger sizes, 
we have checked it for a smaller system, that is, $L=6$.
The estimate of the ground state entropy per spin $S_0/N$ 
for $L=6$ is 0.376.  Thus, the expected value for 
the proportion of the typical BSS ground states 
among all the ground states is 
$6 \cdot 2^{(N/2)}/e^{0.376 N}$ = 0.010 ($N$=216). 
Actually, the typical BSS ground states were visited 
0.010 times as frequently as all the ground states, 
which is consistent with the theoretical expectation. 
In other words, we sample ground states uniformly. 

\begin{figure}[t]
\epsfxsize=\linewidth 
\centerline{\epsfbox{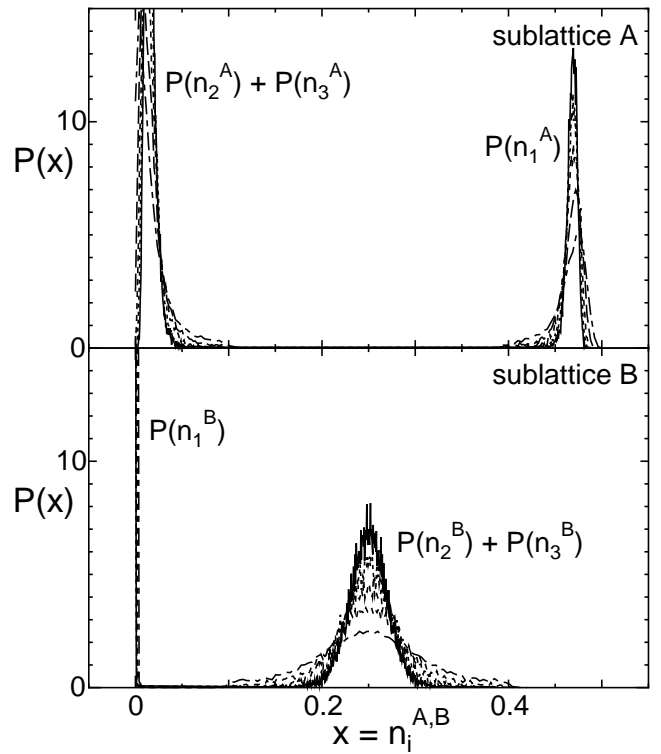}}
\caption{
Distribution function $P(x)$ for the ground states of 
the 3D AF $3$-state Potts models. 
Here $x$ stands for $n_i^{A,B} \ (i=1,2,3)$, 
the number of each component per spin on the sublattice $A$ or $B$.
The linear system sizes are $L$ = 8, 10, 12, 14 and 16. 
The same types of lines for different sizes are used as in 
Fig.~\ref{fig3}.}
\label{fig_gd}
\end{figure}

Next we study the distribution of the ground states. 
As an example, we again deal with the case of $q=3$. 
Let us denote the number of each component per spin as 
$n_i^{A,B} \ (i=1,2,3)$;  then,
\begin{equation}
 0 \le n_i \le 1/2 \quad {\rm and} \quad 
 m_i = n_i^{A}-n_i^{B},
\end{equation}
where the staggered magnetization $m_i$ is given in Eq.~(\ref{order}). 
We show the distribution function of $n_i^{A,B}$ 
for $L$=8, 10, 12, 14 and 16 in Fig.~\ref{fig_gd}.  
The same types of lines for different sizes are used as in 
Fig.~\ref{fig3}.
The distribution is sharper for larger sizes.
In plotting the data in Fig.~\ref{fig_gd}, we have chosen 
the component 1 and the sublattice $A$ such that $n_1^{A}$ 
is the largest.  The distribution function $P(n_i^{A,B})$ 
is normalized such that
$$
\int P(n_i^{A,B}) \ dn_i^{A,B}
$$
is independent of the size.
For the typical BSS states, we expect the $\delta$-function 
distribution for the sublattice $A$, 
$P(n_1^{A}) = \delta(n_1^{A}-1/2)$, 
$P(n_{2,3}^{A}) = \delta(n_{2,3}^{A})$; 
$P(n_{2}^{B})+P(n_{3}^{B})$ becomes the Gaussian distribution 
around 1/4 and $P(n_1^{B}) = \delta(n_1^{B})$
for the sublattice $B$. 
Our results shown in Fig.~\ref{fig_gd} are close to 
those of the typical BSS states, but we can see 
a clear deviation from the typical BSS states and 
the size dependence. 
In other words, there are many configurations other than 
the typical configurations of the BSS state.
Similar behavior is also obtained for 
the ground states of higher-state Potts models. 
The deviation from the typical BSS states becomes larger 
for higher $q$.

\section{Summary and discussions}

To summarize, we have applied a newly proposed Monte Carlo 
algorithm, the Wang-Landau algorithm \cite{wl01}, 
to the study of the 3D AF $q$-state Potts models.
We obtain the finite-temperature phase transition for $q$= 3 and 4, 
whereas the transition temperature is down to zero for $q$=5. 
For $q$=6 there exists no order for all the temperatures. 
We also study the ground-state properties.
From the analysis of the size-dependence of the ground-state 
entropy, we find that the ground-state entropy 
is larger than the contribution from the typical configurations 
of the BSS state for $q=3$. 
The same situations are found for $q$ = 4, 5 and 6. 

We have confirmed again the efficiency of 
the Wang-Landau algorithm.  
For the study of only the critical phenomena 
near the critical point, other methods may have advantage.  
However, in order to make the systematic study for 
all the energy space, we can treat larger systems 
by using the Wang-Landau algorithm,
especially for higher $q$-state Potts models.

In the present paper we have studied the ground-state 
entropy for the 3D AF Potts models. 
We had better mention that the accurate estimate of 
the nonzero ground-state entropy is easily obtained 
for the 2D AF Potts models, of course; the obtained data 
are consistent with the previous studies \cite{Shrock}. 
It is also interesting to apply the Wang-Landau algorithm
to the systematic study of both the ground-state 
properties and the phase transitions for more complicated 
systems, such as spin glass problems.  

\section*{Acknowledgments}

We thank N. Kawashima, H. Otsuka, M. Kikuchi and F. Wang 
for valuable discussions.  
The computation in this work has been done using the facilities of
the Supercomputer Center, Institute for Solid State Physics,
University of Tokyo.
This work was supported by a Grant-in-Aid for Scientific Research 
from the Ministry of Education, Science, Sports and Culture, Japan.


\vspace{1cm}

\vspace{1cm}

\end{multicols}
\end{document}